\documentclass{article}

\usepackage{graphicx}
\usepackage{amssymb}
\usepackage{amsmath}
\usepackage{amscd}
\usepackage{mathrsfs}
\usepackage{latexsym}

\title{\textbf{Extended Representations of Observables and States for a Noncontextual Reinterpretation of QM}}

\author{Claudio Garola\footnote{Department of Physics, University of Salento - Via Arnesano 73100 Lecce, Italy; INFN, Lecce Section - Via Arnesano 73100 Lecce, Italy, EU; email: garola@le.infn.it} \ \ and \ Sandro Sozzo\footnote{Center Leo Apostel (CLEA), Free University of Brussels (VUB) - Krijgskundestraat 33, 1150 Brussels, Belgium; Department of Physics, University of Salento - Via Arnesano 73100 Lecce, Italy; email: ssozzo@vub.ac.be, sozzo@le.infn.it}}

\begin{document}

\maketitle

\begin{abstract}
\noindent
A crucial and problematical feature of quantum mechanics (QM) is nonobjectivity of properties. The \emph{ESR model} restores objectivity reinterpreting quantum probabilities as \emph{conditional} on detection and embodying the mathematical formalism of QM into a broader noncontextual (hence local) framework. We propose here an improved presentation of the ESR model containing a more complete mathematical representation of the basic entities of the model. We also extend the model to mixtures showing that the mathematical representations of \emph{proper} mixtures does not coincide with the mathematical representation of mixtures provided by QM, while the representation of \emph{improper} mixtures does. This feature of the ESR model entails that some interpretative problems raising in QM when dealing with mixtures are avoided. From an empirical point of view the predictions of the ESR model depend on some parameters which may be such that they are very close to the predictions of QM in most cases. But the nonstandard representation of proper mixtures allows us to propose the scheme of an experiment that could check whether the predictions of QM or the predictions of the ESR model are correct.
\end{abstract}

\section{Introduction\label{intro}}
A crucial feature of the standard interpretation of quantum mechanics (QM) is \emph{nonobjectivity} of physical properties, which follows from a series of ``no--go theorems'', the most important of which are the Bell \cite{b64} and Bell--Kochen--Specker \cite{b66,ks67} theorems. To be precise, if one adopts a minimal ``realistic'' position, according to which individual examples of physical systems can be produced \cite{blm91,bs96,b98}, one can supply an operational definition of objectivity by stating that a physical property $E$ (\emph{e.g.}, the value of an observable) is objective for a given state $S$ of a physical system $\Omega$ if for every individual example of $\Omega$ in the state $S$ the result of an ideal measurement of $E$ does not depend on the measurement context. Then, the Bell--Kochen--Specker theorem provides examples of physical systems and states in which there are nonobjective properties (\emph{contextuality} of QM), while the Bell theorem shows that contextuality may occur also at a distance (\emph{nonlocality of QM}), both features supporting the standard assumption in QM that for every state there are physical properties that are not objective.

Nonobjectivity of physical properties has many puzzling consequences. From a logical point of view it implies that the classical notion of truth as correspondence cannot be maintained in QM, hence many authors state that a nonclassical logic (\emph{Quantum Logic}) has to be adopted in the language of QM \cite{bvn36,r98,dcgg04}. From a probabilistic point of view it implies that the usual \emph{epistemic} interpretation of probabilities cannot be maintained in the case of quantum probabilities, which are necessarily nonepistemic (or \emph{ontic}) \cite{blm91,bc81}. From a physical point of view it entails the \emph{objectification problem}, \emph{i.e.}, the main and unsolved problem of the quantum theory of measurement \cite{blm91,bs96,b98}, hence several known paradoxes (Sch\"{o}dinger's cat, Wigner's friend, etc.). From an intuitive point of view, it implies that models which would entail objectivity of physical properties cannot be constructed in QM (wave--particle duality). 

The above consequences of nonobjectivity are intriguing but they also raise many problems, and a huge literature has been devoted to them. We limit ourselves here to quote a famous sentence by Feynman \cite{f65}.

``There was a time when the newspapers said that only twelve men understood the theory of relativity. I do not believe there ever was such a time. There might have been a time when only one man did, because he was the only guy who caught on before he wrote his paper. But after people read the paper a lot of people understood the theory of relativity in some way or other, certainly more than twelve. On the other hand I think I can safely say that nobody understands quantum mechanics.''

Trying to avoid the foregoing problems, we have recently published several papers \cite{ga03,gp04,gs09,gs08,sg08,gs10,gs10b,gs10c} in which an ESR (\emph{extended semantic realism}) model is worked out whose mathematical apparatus embodies the mathematical apparatus of QM but quantum probabilities are reinterpreted as conditional rather than absolute. The ESR model is a noncontextual (hence local) \emph{hidden variables} theory, according to which physical properties are objective and the no--go theorems do not hold because of the aforesaid reinterpretation of quantum probabilities. It consists of a microscopic and a macroscopic part, and the latter can be presented as a self--consistent theory, without mentioning the hidden variables, even if the hidden variables are needed if one has to prove objectivity or to justify the assumptions introduced at a macroscopic level. The new theory yields predictions that are generally different from the predictions of QM, but the difference depends on some parameters (the \emph{detection probabilities}) which may make it so small that it remains unnoticed at the experimental level. There are however physical situations in which it becomes relevant and one can contrive experiments to check which predictions better fit experimental data \cite{gs08,gs10,gs10b}.

We intend to supply a new, more complete version of the ESR model in this paper, in particular providing mathematical representations of improper mixtures and their transformations when (idealized) measurements occur. Therefore we provide an introduction to the ESR model in Sec. \ref{modello}, proposing new mathematical representations of macroscopic properties and generalized observables and reporting from previous papers \cite{gs09,gs10} a \emph{generalized projection postulate} (GPP) which yields the transformation of a pure state induced by an idealized nondestructive measurement. Then we discuss the operational definitions of proper and improper mixtures in Sec. \ref{ignorance}. Basing on these definitions, we resume and complete some previous results in Sec. \ref{gen_obse_mixed_case} which show that the mathematical representation of proper mixtures in the ESR model is different from the mathematical representation of mixtures in QM, and that a \emph{generalized L\"{u}ders postulate} (GLP) can be stated which yields the transformation of a proper mixture induced by an idealized measurement. Moreover we prove in Sec. \ref{repres_improper} that the ESR model implies a mathematical representation of improper mixtures which coincides with the representation of mixtures provided by QM and show that the transformation of improper mixtures induced by an idealized measurement can be obtained from GPP by substituting in it the representation of a pure state with the representation of an improper mixture. We can thus conclude that the ESR model neatly distinguishes proper from improper mixtures and that its probabilistic predictions do not formally coincide with the predictions of QM in the case of proper mixtures. This offers a solution of some problems raised by the standard quantum representation of mixtures and allows us to propose in Sec. \ref{testable} a scheme for an experiment checking whether the predictions of QM or the predictions of the ESR model are correct.

\section{The ESR model\label{modello}}
According to the ESR model, a physical system $\Omega$ is operationally defined by a pair $(\Pi, {\mathscr R})$, with $\Pi$ a set of \emph{preparing devices} and ${\mathscr R}$ a set of \emph{measuring apparatuses}. Every preparing device, when activated, prepares an \emph{individual example} of $\Omega$ (which can be identified with the preparation act itself if one wants to avoid any ontological commitment). Every measuring device, if activated after a preparing device, yields an \emph{outcome}, that we assume to be a real number.

In the theoretical description a physical system $\Omega$ is characterized by a set ${\mathscr U}$ of \emph{physical objects} and a set $\mathcal E$ of \emph{microscopic properties} at a microscopic level, and by a set $\mathcal S$ of \emph{states} and a set ${\mathcal O}_{0}$ of \emph{generalized observables} at a macroscopic level.

Physical objects are operationally interpreted as individual examples of $\Omega$, while microscopic properties are purely theoretical entities (the hidden variables of the model). Every physical object $x \in {\mathscr U}$ is associated with a set of microscopic properties (the microscopic properties \emph{possessed} by $x$) which is called the \emph{microscopic state} of $x$ and also is a theoretical entity.
 
States are physically defined as classes of probabilistically equivalent preparing devices, following standard procedures in the foundations of QM \cite{bc81}. Every device $\pi \in S \in {\mathcal S}$, when constructed and activated, prepares an individual example of $\Omega$, hence a physical object $x$, and one briefly says that ``$x$ is (prepared) in the state $S$''. Analogously, generalized observables are physically defined as classes of probabilistically equivalent \emph{measuring apparatuses}. Every $A_0 \in {\mathcal O}_{0}$ is obtained by considering an observable $A$ in the set ${\mathcal O}$ of all observables of QM and adding a \emph{no--registration outcome} $a_0$ to the set $\Xi$ of all possible values of $A$ on the real line $\Re$, so that the set of all possible values of $A_0$ is $\Xi_0=\{ a_0 \} \cup \Xi$.\footnote{We assume here that $\Re \setminus \Xi$ is non--void, which is not restrictive. Indeed, if $\Xi=\Re$, one can choose a bijective Borel function $f: \Re \rightarrow \Xi'$ such that $\Xi' \subset \Re$, and replace $A$ with $f(A)$. \label{borel_sets}}\footnote{One could obtain a more general theory by introducing \emph{unsharp} observables at this stage. For the sake of simplicity we do not consider this generalization of the ESR model in the present paper.}

The set ${\mathcal F}_{0}$ of all (\emph{macroscopic}) \emph{properties} of $\Omega$ is then defined as follows,
\begin{equation}
{\mathcal F}_{0} \ =  \ \{ (A_0, X) \ | \  A_0 \in {\mathcal O}_{0}, \ X \in \mathbb{B}(\Re) \},
\end{equation}
where $\mathbb{B}(\Re)$ is the $\sigma$--algebra of all Borel subsets of $\Re$. Hence the subset 
\begin{equation}
{\mathcal F} \ =  \ \{ (A_0, X) \ | \  A_0 \in {\mathcal O}_{0}, \ X \in \mathbb{B}(\Re), \ a_0 \notin X \} \subset {\mathcal F}_{0}
\end{equation}
is in one--to--one correspondence with with the set $\{ (A, X) \ | \  A \in {\mathcal O}, \ X \in \mathbb{B}(\Re) \}$ of all properties associated with observables of QM.

A measurement of a property $F=(A_0, X)$ on a physical object $x$ in the state $S$ is then described as a \emph{registration} performed by means of a \emph{dichotomic registering device} whose outcomes are denoted by \emph{yes} and \emph{no}. The measurement yields outcome yes/no (equivalently, $x$ \emph{displays}/\emph{does not display} $F$), if and only if the value of $A_0$ belongs/does not belong to $X$.

The connection between the microscopic and the macroscopic part of the ESR model is established by introducing the following assumptions.

(i) A bijective mapping $\varphi: {\mathcal E} \longrightarrow {\mathcal F} \subset {\mathcal F}_{0}$ exists.

(ii) If a physical object $x$ is in the microscopic state $s$ and an \emph{idealized measurement} of a macroscopic property $F=\varphi (f)$ is performed on $x$, then $s$ determines a probability $p_{s}^{d}(F)$ that $x$ be detected, and $x$ displays $F$ if it is detected and $f \in s$, does not display $F$ if it is not detected or $f \notin s$. For the sake of simplicity, we will consider only idealized measurements (simply called measurements from now on) in the following.

The ESR model is \emph{deterministic} if $p_{s}^{d}(F) \in \{0,1 \}$, \emph{probabilistic} otherwise. In the former case it is necessarily noncontextual, hence physical properties are objective, because the outcome of the measurement of a macroscopic property on a physical object $x$ depends only on the microscopic properties possessed by $x$ and not on the measurement context. In the latter case one can recover noncontextuality by adding further hidden variables which make $p_{S}^{d}(F)$ epistemic to the microscopic properties \cite{gs08}.\footnote{The idealized measurements introduced in the ESR model correspond to the \emph{ideal measurements} of QM, and noncontextuality refers to idealized measurements only (local contextuality can indeed occur in actual measurements, \emph{e.g.}, whenever a threshold exists \cite{a2009}).}
  
By using the connection between the microscopic and the macroscopic part of the ESR model one can show \cite{gs08} that, whenever the property $F=(A_0, X) \in \mathcal F$ (hence $a_0 \notin X$) is measured on a physical object $x$ in the macroscopic state $S$, the overall probability $p_{S}^{t}(F)$ that $x$ display $F$ is given by
\begin{equation} \label{formuladipartenza}
p_{S}^{t}(F)= p_{S}^{d}(F)p_{S}(F) \ .
\end{equation}

The symbol $p_{S}^{d}(F)$ in Eq. (\ref{formuladipartenza}) denotes the probability that $x$ be detected whenever it is in the state $S$ (\emph{detection probability}) and $F$ is measured. The value of $p_{S}^{d}(F)$ is not fixed for a given generalized observable $A_0$ because it may depend on $F$, hence on $X$. But the connection of microscopic with macroscopic properties via $\varphi$ implies that $p_{S}^{d}(F)$ depends only on the features of the physical objects in the state $S$, hence it does not occur because of flaws or lack of efficiency of the apparatus measuring $F$. 

The symbol $p_{S}(F)$ in Eq. (\ref{formuladipartenza}) denotes instead the conditional probability that $x$ display $F$ when it is detected.

Let us consider now the measurement of a property $F=(A_0, X)\in {\mathcal F}_{0} \setminus {\mathcal F}$ (hence $a_0 \in X$). Putting $F^{c}=(A_0, \Re \setminus X)$, we introduce the physically reasonable assumption that, for every state $S$,
\begin{equation} \label{f_bar_c}
p_{S}^{t}(F)=1-p_{S}^{t}(F^{c})=1-p_{S}^{d}(F^{c})p_{S}(F^{c}).
\end{equation}
Eq. (\ref{f_bar_c}) provides the overall probability $p_{S}^{t}(F)$ that a physical object $x$ in the state $S$ display $F$ in terms of the overall probability that $x$ display $F^{c}$ when $F^{c}$ is measured in place of $F$. 

Coming back to a property $F \in {\mathcal F}$, Eq. (\ref{formuladipartenza}) implies that three basic probabilities occur in the ESR model. We have as yet no theory which allows us to predict the value of $p_{S}^{d}(F)$. But we can consider $p_{S}^{d}(F)$ as an unknown parameter to be determined empirically and introduce theoretical assumptions that connect the ESR model with QM enabling us to provide  mathematical representations of the physical entities introduced in the ESR model together with explicit expressions of $p_{S}^{t}(F)$ and $p_{S}(F)$.

Let us begin with $p_{S}(F)$. Then, the following statement expresses the fundamental assumption of the ESR model.

\vspace{.1cm}
\noindent
\emph{AX. If $S$ is a pure state the probability $p_{S}(F)$ can be evaluated by using the same rules that yield the probability of the property $F$ in the state $S$ according to QM.}

\vspace{.1cm} 
Assumption AX allows one to recover the basic formalism of QM in the framework of the ESR model but modifies its standard interpretation. Indeed, according to QM, whenever an ensemble ${\mathscr E}_{S}$ of physical objects in a state $S$ is prepared and ideal measurements of a property $F$ are performed, all physical objects in ${\mathscr E}_{S}$ are detected, hence the quantum rules yield the probability that a physical object $x$ display $F$ if $x$ is selected in ${\mathscr E}_{S}$ (\emph{absolute} probability). According to assumption AX, instead, if $S$ is pure, the quantum rules yield the probability that a physical object $x$ display $F$ if idealized measurements of $F$ are performed and $x$ is selected in the subset of all objects of ${\mathscr E}_{S}$ that are detected (\emph{conditional} probability).

Because of the above reinterpretation of quantum probabilities the predictions of the ESR model are different from those of QM. As we have anticipated in Sec. \ref{intro}, the difference depends on the detection probabilities and may be very small (because $p_{S}^{t}(F)$ is close to $p_{S}(F)$ if $p_{S}^{d}(F)$ is close to 1), so that it remains unnoticed. Moreover it is hard to distinguish the detection probabilities from the efficiencies of actual measuring devices, which also explains why QM ignores them. Nevertheless we will show in the following that in some cases there are substantial differences between the two theories that can be experimentally checked.

Let us discuss now the mathematical representations of the physical entities introduced in the ESR model. 

Let $F=(A_0, X) \in {\mathcal F}$ (hence $a_0 \notin X$) and let us firstly consider the conditional probability $p_{S}(F)$. Let $S$ be a pure state. Then assumption AX implies that, as far as $p_{S}(F)$ is concerned, $S$ can be represented by a vector $|\psi\rangle$ in the set ${\mathscr V}$ of all unit vectors of the (separable) complex Hilbert space ${\mathscr H}$ associated with $\Omega$, or by the one--dimensional projection operator $\rho_{\psi}=|\psi\rangle\langle\psi|$, as in QM. Moreover $F$ can be represented by the projection operator $P^{\widehat{A}}(X)$, where $P^{\widehat{A}}$ is the spectral projection valued (PV) measure associated with the self--adjoint operator $\widehat{A}$ which represents the observable $A \in {\mathcal O}$ of QM from which $A_0$ is obtained. Finally, the probability $p_{S}(F)$ can be calculated by using the standard quantum rule
\begin{equation} \label{prob_X_QM}
p_{S}(F)=\langle\psi|P^{\widehat{A}}(X)|\psi\rangle=Tr [\rho_{\psi}P^{\widehat{A}}(X)] ,
\end{equation}
hence also the generalized observable $A_0 \in {\mathcal O}_{0}$ can be represented by $\widehat{A}$.

Let us come to the overall probability $p_{S}^{t}(F)$. Let $d \lambda$ be an interval on $\Re$ with $a_0 \notin d\lambda$, and let $dp_{S}^{t}$ be the overall probability that a measurement of $A_0$ on a physical object $x$ in the pure state $S$ considered above yield an outcome in $d \lambda$. Then, Eq. (\ref{formuladipartenza}) and assumption AX suggest that
\begin{equation}
dp_{S}^{t}={p}_{\psi}^{d}(\widehat{A}, \lambda) \langle\psi|P^{\widehat{A}}(d \lambda)|\psi\rangle
\end{equation}
where ${p}_{\psi}^{d}(\widehat{A}, \lambda)$ is a detection probability such that the function $\langle\psi| {p}_{\psi}^{d}(\widehat{A}, \lambda)\frac{P^{\widehat{A}}(d \lambda)}{d \lambda}|\psi\rangle$ is measurable on $\Re$. It follows
\begin{equation} \label{prob_X_W}
p_{S}^{t}((A_0, X))=p_{S}^{t}(F)=\langle\psi|T_{\psi}^{\widehat{A}}(X)|\psi\rangle=Tr [\rho_{\psi}T_{\psi}^{\widehat{A}}(X)]
\end{equation}
with
\begin{equation} \label{POV_lebesgue}
T_{\psi}^{\widehat{A}}(X) =\int_{X}{p}_{\psi}^{d}(\widehat{A}, \lambda) P^{\widehat{A}}( \mathrm{d}\lambda) \quad (a_0 \notin X) .
\end{equation}
Therefore, as far as $p_{S}^{t}(F)$ is concerned, the pure state $S$ can still be represented by $|\psi\rangle$ or $\rho_{\psi}$. The property $F$ is represented instead by the family $\{ T_{\psi}^{\widehat{A}}(X) \}_{|\psi\rangle\in {\mathscr V}}$. 

Putting together the results obtained above, we conclude that the mathematical representation of a property $F=(A_0, X) \in {\mathcal F}$ is provided by the pair
\begin{equation}
(P^{\widehat{A}}(X), \{ T_{\psi}^{\widehat{A}}(X) \}_{|\psi\rangle\in {\mathscr V}}) ,
\end{equation}
where the first element of the pair coincides with the representation in QM of the property $(A, X)$ and must be used to calculate $p_{S}(F)$, while the second element must be used to calculate $p_{S}^{t}(F)$.

Let us consider now a property $F=(A_0, X) \in {\mathcal F}_{0} \setminus {\mathcal F}$ (hence $a_0 \in X$). By using Eqs. (\ref{f_bar_c}), (\ref{prob_X_W}) and (\ref{POV_lebesgue}) we get
\begin{equation}
p_{S}^{t}(F)=1-\langle\psi|T_{\psi}^{\widehat{A}}(\Re \setminus X)|\psi\rangle=\langle\psi|(I-\int_{\Re \setminus X}{p}_{\psi}^{d}(\widehat{A}, \lambda) P^{\widehat{A}}( \mathrm{d}\lambda))|\psi\rangle
\end{equation}
where $I$ is the identity operator on $\mathscr H$. Hence Eq. (\ref{prob_X_W}) still holds if $F \in {\mathcal F}_{0} \setminus {\mathcal F}$ and
\begin{equation} \label{POV_lebesgue_0}
T_{\psi}^{\widehat{A}}(X) =I-\int_{\Re \setminus X}{p}_{\psi}^{d}(\widehat{A}, \lambda) P^{\widehat{A}}( \mathrm{d}\lambda) \quad (a_0 \in X) .
\end{equation}
For every $|\psi\rangle\in {\mathscr V}$ we can thus introduce a positive operator valued (POV) measure 
\begin{equation} \label{math_rep_gen_obs}
T_{\psi}^{\widehat{A}}: X \in \mathbb{B}(\Re) \longmapsto T_{\psi}^{\widehat{A}}(X) \in {\mathscr B}({\mathscr H}) ,
\end{equation}
where ${\mathscr B}({\mathscr H})$ is the set of all bounded operators on $\mathscr H$, defined by Eqs. (\ref{POV_lebesgue}) and (\ref{POV_lebesgue_0}). This measure is commutative, that is, for every $X, Y \in \mathbb{B}(\Re), T_{\psi}^{\widehat{A}}(X)T_{\psi}^{\widehat{A}}(Y)=T_{\psi}^{\widehat{A}}(Y)T_{\psi}^{\widehat{A}}(X)$. Moreover, for every pure state $S$ represented by the vector $|\psi\rangle\in {\mathscr V}$ and every Borel set $X$, the family
\begin{equation} \label{family_gen_bs}
{\mathcal T}^{\widehat{A}}=\left \{ T_{\psi}^{\widehat{A}}\right \}_{|\psi\rangle\in {\mathscr V}}
\end{equation}
provides the overall probability that the outcome of a measurement of $A_0$ on a physical object $x$ in the state $S$ belong to the Borel set $X$ via Eq. (\ref{prob_X_W}). 

Putting together the results obtained above, we conclude that the mathematical representation of a generalized observable $A_0 \in {\mathcal O}_{0}$ is provided by the pair
\begin{equation}
(\widehat{A},{\mathcal T}^{\widehat{A}}) ,
\end{equation}
where the first element of the pair coincides with the representation in QM of the observable $A$.

Finally, one gets from assumption AX, Eq. (\ref{formuladipartenza}) and Eq. (\ref{prob_X_W}) that, for every $|\psi\rangle \in {\mathscr V}$ and $F=(A_0, X)\in \mathcal F$, 
\begin{equation} \label{useful_relation}
p_{S}^{d}(F)=\frac{Tr[\rho_{\psi}T_{\psi}^{\widehat{A}}(X)]}{Tr[\rho_{\psi} P^{\widehat{A}}(X)]} \ ,
\end{equation}
which yields a condition that must be fulfilled by $p_{S}^{d}(F)$.

We conclude this section by resuming the \emph{generalized projection postulate} presented in some recent papers \cite{gs09,gs10,gs10c}.

\vspace{.1cm}
\noindent
\emph{GPP}. \emph{Let $S$ be a pure state represented by the density operator $\rho_{\psi}$, and let a nondestructive idealized measurement of the physical property $F=(A_0,X)\in {\mathcal F}_{0}$ be performed on a physical object $x$ in the state $S$.} 

\emph{Let the measurement yield the yes outcome. Then, the state $S_F$ of $x$ after the measurement is a pure state represented by the density operator}
\begin{equation} \label{genpost_dis_W}
\rho_{\psi_{F}}=\frac{T_{\psi}^{\widehat{A}}(X)\rho_{\psi}T_{\psi}^{\widehat{A} \dag}(X)}{Tr[T_{\psi}^{\widehat{A}}(X)\rho_{\psi}T_{\psi}^{\widehat{A} \dag}(X)]} \ .
\end{equation}

\emph{Let the measurement yield the no outcome. Then, the state $S'_{F}$ of $x$ after the measurement is a pure state represented by the density operator $\rho_{\psi'_{F}}$ obtained by replacing $T_{\psi}^{\widehat{A}}(X)$ with $T_{\psi}^{\widehat{A}}(\Re \setminus X)$ in the second member of Eq. (\ref{genpost_dis_W})}

\vspace{.1cm}

GPP generalizes the L\"{u}ders postulate of QM in the case of pure states. It does not imply, however, any objectification because all physical properties are objective in the ESR model, as we have seen above.

\section{The operational definitions of proper and improper mixtures\label{ignorance}}
It is well known that some authors distinguish two kinds of mixtures in QM, that is, \emph{proper} and \emph{improper} mixtures \cite{blm91,des76,tb05}. Other authors instead argue that only improper mixtures exist according to QM \cite{bc81,f57,p68,ba98}. From a mathematical point of view all mixtures are represented by density operators in QM, hence no distinction occurs. Nevertheless there are basic differences between the physical definitions of the two kinds of mixtures which entail that their mathematical representations do not coincide in the ESR model. Let us therefore discuss this topic in more details.

(i) \emph{Proper mixtures}. Bearing in mind the physical definition of states mentioned in Sec. \ref{modello}, physical objects in a pure state $S$ can be prepared by activating repeatedly any preparing device $\pi$ in the equivalence class defining $S$. Moreover, the preparation procedure of a physical object $x$ in a state $M$ which is a proper mixture of the pure states $S_1, S_2, \ldots$ with probabilities $p_1, p_2, \ldots$, respectively, can be described as follows.

\emph{Choose a preparing device $\pi_{j}$ for every pure state $S_{j}$, use each $\pi_j$ to prepare an ensemble ${\mathscr E}_{S_j}$ of $n_j$ physical objects in the pure state $S_j$ and choose $n_j$ such that $n_j=n p_j$, with $n=\sum_{j}n_j$. Then mingle the ensembles ${\mathscr E}_{S_1}, {\mathscr E}_{S_2}, \ldots$ to prepare an ensemble ${\mathscr E}_{M}$ of $n$ physical objects, remove any memory of the way in which ${\mathscr E}_{S_1}, {\mathscr E}_{S_2}, \ldots$ have been mingled and select a physical object in ${\mathscr E}_{M}$}.

The above description implies that one assumes that frequencies converge to probabilities in the large number limit and interprets the  probabilities $p_1, p_2, \ldots$ as epistemic, \emph{i.e.}, formalizing the loss of memory about the pure state in which each physical object has been prepared (\emph{ignorance interpretation}). It also implies that many preparation procedures of a physical object in a state $M$ can be constructed, hence we call \emph{operational definition} of $M$ and denote by $\sigma_M$ the set of preparation procedures of $M$ that can be obtained by selecting the preparing devices in the pure states $S_1$, $S_2, \ldots$ in all possible ways.

Let us recall now that the state $M$ is represented in QM by a density operator $\rho_{M}=\sum_{j}p_j \rho_{\psi_{j}}$, where $\rho_{\psi_{j}}=|\psi_{j}\rangle\langle\psi_{j}|$ is the density operator representing the pure state $S_{j}$. It is then well known that one--dimensional projection operators $\rho_{\chi_1}, \rho_{\chi_2}, \ldots$ generally exist, none of which coincides with one of the projection operators $\rho_{\psi_1}, \rho_{\psi_2}, \ldots$, which are such that $\rho_M=\sum_{l} q_l \rho_{\chi_l}$, with $0\le q_l \le 1$ and $\sum_{l} q_l=1$. If this expression of $\rho_M$ is adopted, the coefficients $q_l$ cannot be interpreted as probabilities bearing an ignorance interpretation because $\rho_{\chi_l}$ does not represent a possible pure state of a physical object in the state $M$.  Consider then a proper mixture $M'$ of the pure states $T_1, T_2, \ldots$ represented by the density operators $\rho_{\chi_1}, \rho_{\chi_2}, \ldots$, with probabilities $q_1, q_2, \ldots$, respectively. This mixture has an operational definition $\sigma_{M'}$ which is different from $\sigma_M$. Notwithstanding this, the choice of probabilities and states implies that $\rho_M=\rho_{M'}$, hence $\sigma_M$ and $\sigma_{M'}$ are probabilistically equivalent and $M'$ must be identified with $M$ according to QM. But the probabilities $q_1, q_2, \ldots$ now admit an ignorance interpretation, at variance with the conclusion obtained when $M$ is considered. It follows that some physical information supplied by the operational definition of $M$ goes lost when $M$ is represented by $\rho_{M}$, and that the ambiguities in the interpretation of the mathematical formalism, widely debated in the literature, can be ascribed to this loss.

(ii) \emph{Improper mixtures}. We have considered so far physical objects that are prepared by activating preparing devices that produce examples of a given physical system $\Omega$. But examples of $\Omega$ can also be obtained by preparing examples of a composite physical system $\Gamma$ such that $\Omega$ is a subsystem of $\Gamma$. A typical preparation procedure of this kind can be described as follows. 

\emph{Consider a composite physical system $\Gamma$ made up of two subsystems $\Omega$ and $\Delta$. Choose a preparing device $\pi \in S$, with $S$ a pure state of $\Gamma$, prepare a set ${\mathscr E}_{S}$ of individual examples of $\Gamma$, select an element of ${\mathscr E}_{S}$ and consider the part $x$ of it that constitutes an individual example of $\Omega$}.

When the above procedure is applied one can attribute a state $N$ to $x$ which is represented in QM by a density operator $\rho_{N}$ obtained by tracing over $\Delta$ the one--dimensional projection operator representing $S$. If $\rho_{N}$ also is a one--dimensional projection operator, $N$ is considered as a pure state in a standard sense. Otherwise $N$ is said to be an \emph{improper mixture}. In the latter case the preparation procedure does not privilege any convex decomposition of $\rho_{N}$ into one--dimensional projection operators representing pure states of $\Omega$, hence $N$ can be considered as a mixture of pure states in (infinitely) many different ways, with coefficients that can be interpreted as probabilities but never admit an ignorance interpretation.

We note that the distinction between proper and improper mixtures is often overlooked by physicists because the two kinds of mixtures have the same mathematical representations in QM. But the preparation procedures described above imply that proper and improper mixtures are empirically disinguishable, as already stressed by some authors \cite{tb05}. Indeed, if one prepares an ensemble ${\mathscr E}_{N}$ of physical objects in the improper mixture $N$, every subensemble of ${\mathscr E}_{N}$ has the same statistical properties possessed by ${\mathscr E}_{N}$ (that is, it is a fair sample of ${\mathscr E}_{N}$), which does not occur if one prepares an ensemble ${\mathscr E}_{M}$ of physical objects in the proper mixture $M$.

\section{The representation of proper mixtures in the ESR model\label{gen_obse_mixed_case}}
We have recently provided a mathematical representation of proper mixtures in the ESR model \cite{gs10,gs10c}. We resume and complete it here as follows.

Let $M$ be a proper mixture of the pure states $S_1, S_2, \ldots$, represented by the density operators $\rho_{\psi_1}$, $\rho_{\psi_2}$, \ldots, with probabilities $p_1, p_2, \ldots$, respectively. The probability $p_{M}^{t}((A_0, X))$ that a measurement of the generalized observable $A_0$ on a physical object $x$ in the state $M$ yield an outcome in the Borel set $X \in \mathbb{B}(\Re)$, with $a_0 \notin X$, or, equivalently, the probability $p_{M}^{t}(F)$ that $x$ display the macroscopic property $F=(A_0, X) \in \mathcal F$ when $F$ is measured on it, is given by
\begin{equation} \label{prob_tot_mix_phys}
p_{M}^{t}(F)=\sum_{j} p_{j} p_{S_j}^{t}(F)=\sum_{j} p_j p_{S_j}^{d}(F) p_{S_j}(F),
\end{equation}
because of Eq. (\ref{formuladipartenza}), with $S_j$ in place of $S$. By using again Eq. (\ref{formuladipartenza}) with $M$ in place of $S$ we also get
\begin{equation} \label{prob_mix_phys}
p_{M}(F)=\sum_{j} p_{j} \frac{p_{S_j}^{d}(F)}{p_{M}^{d}(F)} p_{S_j}(F).
\end{equation}
Eq. (\ref{prob_mix_phys}) is intuitively reasonable. Indeed, the factor $p_j \frac{p_{S_j}^{d}(F)}{p_{M}^{d}(F)}$ can be interpreted, because of the Bayes law, as the conditional probability that $x$ be in the state $S_j$ whenever $F$ is measured and $x$ is detected. 

Eq. (\ref{prob_X_QM}) entails $p_{S_j}(F)=Tr [\rho_{\psi_j} P^{\widehat{A}}(X)]$. Hence we get from Eq. (\ref{prob_mix_phys}) 
\begin{equation} \label{AX_mixed_0}
p_{M}(F)=Tr [\rho_M(F) P^{\widehat{A}}(X)]
\end{equation}
with
\begin{equation} \label{rho_esse_effe}
\rho_{M}(F)=\sum_{j} p_j \frac{p_{S_j}^{d}(F)}{p_{M}^{d}(F)} \rho_{\psi_j}  \ .
\end{equation}
Furthermore, if we introduce the obvious assumption
\begin{equation} \label{prob_mix_d}
p_{M}^{d}(F)=\sum_{j} p_j p_{S_j}^{d}(F)
\end{equation}
and use Eq. (\ref{useful_relation}) with $S_j$ and $\psi_j$ in place of $S$ and $\psi$, respectively, we get
\begin{equation} \label{rho_S(F)}
\rho_{M}(F)= \frac{\sum_{j} p_j
\frac{Tr[\rho_{\psi_j}T_{\psi_j}^{\widehat{A}}(X)]}{Tr[\rho_{\psi_j} P^{\widehat{A}}(X)]} \rho_{\psi_j}}{\sum_{j} p_j \frac{Tr[\rho_{\psi_j}T_{\psi_j}^{\widehat{A}}(X)]}{Tr[\rho_{\psi_j} P^{\widehat{A}}(X)]}}  \ .
\end{equation}
Eqs. (\ref{AX_mixed_0}) and (\ref{rho_S(F)}) show that $p_{M}(F)$ does not coincide, in general, with the probability obtained by applying standard QM rules, \emph{i.e.}, calculating $Tr[\rho_{M} P^{\widehat{A}}(X)]$, where $\rho_{M}=\sum_{j} p_j \rho_{\psi_{j}}$ is the density operator that represents $M$ in QM. This can be intuitively explained by observing that an ensemble ${\mathscr E}_{M}$ of physical objects prepared in $M$ can be partitioned into subensembles ${\mathscr E}_{S_1}, {\mathscr E}_{S_2}, \ldots$ of physical objects prepared in the states $S_1, S_2, \ldots$, respectively. Whenever a macroscopic property $F$ is measured on ${\mathscr E}_{M}$, for every ${\mathscr E}_{S_j}$ the subensemble ${\mathscr E}_{S_j}^{d}$ of detected objects depends not only on $F$ but also on $S_j$ hence, generally, the subensemble ${\mathscr E}_{M}^{d}$ of all detected objects is not a fair sample of ${\mathscr E}_{M}$. As far as $p_{M}(F)$ is concerned, $M$ must then be represented by the density operator $\rho_{M}(F)$, which depends on $F$ and coincides with $\rho_{M}$ only in special cases. 

Let us come back to $p_{M}^{t}(F)$. By using Eq. (\ref{prob_X_W}) we get
\begin{equation}
p_{S_j}^{t}(F)=Tr[\rho_{\psi_j}T_{\psi_j}^{\widehat{A}}(X)],
\end{equation} 
where $T_{\psi_j}^{\widehat{A}}(X)=\int_{X}{p}_{\psi_j}^{d}(\widehat{A}, \lambda)  P^{\widehat{A}}({\mathrm{d}\lambda})$ because of Eq. (\ref{POV_lebesgue}). Hence Eq. (\ref{prob_tot_mix_phys}) yields
\begin{equation} 
p_{M}^{t}(F)=Tr \Big [\sum_{j} p_j \rho_{\psi_j} T_{\psi_j}^{\widehat{A}}(X) \Big ]
\end{equation}
or equivalently, because of Eqs. (\ref{formuladipartenza}) and (\ref{prob_tot_mix_phys}),
\begin{equation}
p_{M}^{t}(F)=Tr [p_{M}^{d}(F)\rho_{M}(F)P^{\widehat{A}}(X)]. \label{prob_tot_mix_math_notin}
\end{equation}


Eqs. (\ref{AX_mixed_0}) and (\ref{prob_tot_mix_math_notin}) imply that, for every $F \in \mathcal F$, one needs $\rho_{M}(F)$ to calculate $p_{M}(F)$ and both $p_{M}^{d}(F)$ and $\rho_{M}(F)$ to calculate $p_{M}^{t}(F)$. Hence a complete mathematical representation of $M$ is provided in the ESR model by the family of pairs
\begin{equation} 
\{ ( \rho_{M}(F), p_{M}^{d}(F) ) \}_{H \in {\mathcal F}}
=\Big \{  \Big (  \frac{\sum_{j} p_j p_{S_j}^{d}(F) \rho_{\psi_j}}{\sum_j p_j p_{S_j}^{d}(F)}, \sum_j p_j p_{S_j}^{d}(F) \Big ) \Big \}_{F \in {\mathcal F}} . \label{family_mixtures}
\end{equation} 
The representation in Eq. (\ref{family_mixtures}) depends on the operational definition $\sigma_M$ of $M$ through the coefficients $p_{S_j}^{d}(F)$, hence a state $M'$ such that $\sigma_{M'} \ne \sigma_{M}$ is generally different from $M$ (in the sense that $M$ and $M'$ lead to different probabilistic predictions). One can then assume that physics is such that $M'$ is necessarily different from $M$, so that operational definitions and mathematical representations of proper mixtures are in one--to--one correspondence \cite{gs10}. Thus, no physical information is lost, and the ambiguities that occur in QM disappear.

In addition, Eq. (\ref{AX_mixed_0}) and (\ref{prob_tot_mix_math_notin}) imply that the probabilistic predictions of the ESR model in the case of proper mixtures are also formally different from the predictions of QM (hence assumption AX cannot be extended to proper mixtures). At least in principle, one can decide experimentally which theory provides correct predictions by devising suitable experiments (see Sec. \ref{testable}).

We have seen in Sec. \ref{modello} that GPP rules the transformation of a pure state induced by an idealized nondestructive measurement. We have recently shown that the results resumed in Sec. \ref{modello}, together with GPP, allow us to predict the state transformation induced by a measurement of the same kind if the state of the measured object is a mixture (\emph{generalized L\"{u}ders postulate}, or GLP). For the sake of completeness, let us report this result here \cite{gs09,gs10}.

\vspace{.1cm}
\noindent
\emph{GLP}. \emph{Let $M$ be a mixture of the pure states $S_1, S_2, \ldots$, represented by the density operators $\rho_{\psi_1}$, $\rho_{\psi_2}$, \ldots, with probabilities $p_1, p_2, \ldots$, respectively, and let a nondestructive idealized measurement of a macroscopic property $F=(A_0,X)\in {\mathcal F}_{0}$ be performed on a physical object $x$ in the state $M$}.

\emph{Let the measurement yield the yes outcome. Then, the state $M_F$ of $x$ after the measurement is a mixture of the pure states $S_{1F}, S_{2F}, \ldots$ represented by the density operators $\rho_{\psi_{1F}}, \rho_{\psi_{2F}}, \ldots$ with probabilities $p_{1F}$, $p_{2F}, \ldots$, respectively, where}
\begin{equation} \label{GLP'_yes}
\rho_{\psi_{j F}}=\frac{T_{\psi_j}^{\widehat{A}}(X)\rho_{\psi_j}T_{\psi_{j}}^{\widehat{A} \dag}(X)}{Tr[T_{\psi_j}^{\widehat{A}}(X) \rho_{\psi_j} T_{\psi_j}^{\widehat{A} \dag}(X)]} \qquad \qquad (j=1,2,\ldots)
\end{equation}
\emph{and}
\begin{equation} \label{prob_bayes_yes}
p_{j F}=p_j \frac{p_{S_j}^{t}((A_0, X))}{p_{S}^{t}((A_0, X))}= p_{j} \frac{Tr[\rho_{\psi_j} T_{\psi_j}^{\widehat{A}}(X)]}{Tr \Big [\sum_{j} p_{j} \rho_{\psi_j} T_{\psi_j}^{\widehat{A}}(X) \Big ]} \qquad  (j=1,2,\ldots).
\end{equation}
\emph{Hence $M_F$ is represented by the family of pairs} 
\begin{eqnarray} 
\{ ( \rho_{M_F}(H), p_{M_{F}}^{d}(H) ) \}_{H \in {\mathcal F}} \nonumber \\ 
= \{ (\sum_{j} p_{j F} \frac{p_{S_{j F}}^{d}(H)}{p_{M_F}^{d}(H)} \rho_{\psi_{j F}}, \sum_{j}p_{j F} p_{S_{j F}}^{d}(H) ) \}_{H \in {\mathcal F}} \ . \label{family_GLP_yes}
\end{eqnarray}

\emph{Let the measurement yield the no outcome. Then, the state $M'_{F}$ of $x$ after the measurement is a mixture of the pure states $S'_{1F}, S'_{2F}, \ldots$ represented by the density operators $\rho_{\psi'_{1F}}, \rho_{\psi'_{2F}}, \ldots$ with probabilities $p'_{1F}$, $p'_{2F}, \ldots$,  respectively, where $\rho_{\psi'_{jF}}$ and $p'_{jF}$ are obtained by replacing $T_{\psi_j}^{\widehat{A}}(X)$ with $T_{\psi_j}^{\widehat{A}}(\Re \setminus X)$ in the second members of Eqs. (\ref{GLP'_yes})
and (\ref{prob_bayes_yes}), respectively.}
\emph{Hence $M'_{F}$ is represented by the family of pairs} 
\begin{eqnarray} 
\{ ( \rho_{M'_F}(H), p_{M'_{F}}^{d}(H) ) \}_{H \in {\mathcal F}} \nonumber \\ 
= \{ (\sum_{j} p'_{j F} \frac{p_{S'_{j F}}^{d}(H)}{p_{M'_F}^{d}(H)} \rho_{\psi'_{j F}}, \sum_{j}p'_{j F} p_{S'_{j F}}^{d}(H) ) \}_{H \in {\mathcal F}} \ . \label{family_GLP_no}
\end{eqnarray}

\vspace{.1cm}

It is apparent that GLP generalizes GPP because Eq. (\ref{GLP'_yes}) coincides with Eq. (\ref{genpost_dis_W}) if $S$ is a pure state.

\section{The representation of improper mixtures in the ESR model\label{repres_improper}}
Let $\Gamma$ be a composite physical system made up of the subsystems $\Omega$ and $\Delta$ associated with the Hilbert spaces ${\mathscr H}$ and ${\mathscr G}$, respectively, so that $\Gamma$ is associated with the Hilbert space ${\mathscr H} \otimes {\mathscr G}$. Let $S$ be a pure state of $\Gamma$ represented by the unit vector $|\Psi\rangle$, or by the projection operator $\rho_{\Psi}=|\Psi\rangle\langle\Psi|$, and let $F=(A_{0},X) \in {\mathcal F}$ be a property of $\Omega$ represented by the pair $(P^{\widehat{A}}(X), \{ T_{\psi}^{\widehat{A}}(X) \}_{|\psi\rangle\in {\mathscr V}})$ (Sec. \ref{modello}). Then $F$ obviously corresponds to a property $\tilde{F}$ of $\Gamma$, and one gets from assumption AX
\begin{equation} \label{qm_composite}
p_{S}(\tilde{F})=Tr [ \rho_{\Psi} P^{\widehat{A}}(X) \otimes I_{\mathscr G}],
\end{equation}
where $I_{\mathscr G}$ is the identity operator on ${\mathscr G}$. Hence standard calculations show that
\begin{equation} \label{qm_partial_trace}
p_{S}(\tilde{F})=Tr [ \rho_{N} P^{\widehat{A}}(X)]
\end{equation}
with $\rho_N$ a density operator obtained by tracing $\rho_{\Psi}$ over $\Delta$. Moreover Eq. (\ref{prob_X_W}) yields
\begin{equation} \label{p_esse_t_f_tilde}
p_{S}^{t}(\tilde{F})=Tr [\rho_{\Psi}T_{\Psi}^{\widehat{\tilde{A}}}(X)]
\end{equation}
where
\begin{equation}
T_{\Psi}^{\widehat{\tilde{A}}}(X) =  T_{\Psi}^{\widehat{A}\otimes I_{\mathscr G}}(X)=
\int_{X}{p}_{\Psi}^{d}(\widehat{A} \otimes I_{\mathscr G}, \lambda) 
 P^{\widehat{A} \otimes I_{\mathscr G}}(\mathrm{d} \lambda) .
\end{equation}
Let us put now ${p}_{\Psi}^{d}(\widehat{A} \otimes I_{\mathscr G}, \lambda)={p}_{\rho_{N}}^{d}(\widehat{A}, \lambda)$ and $T_{\rho_{N}}^{\widehat{A}}(X)=\int_X {p}_{\rho_{N}}^{d}(\widehat{A}, \lambda)  P^{\widehat{A}}(\mathrm{d} \lambda)$. It follows
\begin{equation}
T_{\Psi}^{\widehat{\tilde{A}}}(X)=\int_X {p}_{\rho_{N}}^{d}(\widehat{A}, \lambda) (P^{\widehat{A}}( \mathrm{d} \lambda) \otimes I_{\mathscr G})= T_{\rho_{N}}^{\widehat{A}}(X)\otimes I_{\mathscr G},
\end{equation}
hence, substituting in Eq. (\ref{p_esse_t_f_tilde}) and tracing over $\Delta$,
\begin{equation} \label{new_partial_trace}
p_{S}^{t}(\tilde{F})=Tr [ \rho_{N} T_{\rho_{N}}^{\widehat{A}}(X)] .
\end{equation}
Bearing in mind the operational definition of improper mixtures one can now consider the preparation of an example of $\Gamma$ as a preparation of an example of $\Omega$ in a state $N$ which is an improper mixture, and put $p_{S}(\tilde{F})=p_{N}(F)$, $p_{S}^{t}(\tilde{F})=p_{N}^{t}(F)$. Eqs. (\ref{qm_partial_trace}) and (\ref{new_partial_trace}) then yield
\begin{equation} \label{overall_improper}
p_{N}(F)=Tr [ \rho_{N} P^{\widehat{A}}(X)]
\end{equation}
and 
\begin{equation} \label{conditional_improper}
p_{N}^{t}(F)=Tr [ \rho_{N} T_{\rho_{N}}^{\widehat{A}}(X)] ,
\end{equation}
which show that the density operator $\rho_N$ provides the mathematical representation of $N$. This representation coincides with the standard representation of $N$ in QM and is basically different from the representation of a proper mixture, which shows that the ESR model neatly distinguishes proper from improper mixtures.

We observe now that Eqs. (\ref{overall_improper}) and (\ref{conditional_improper}) can be obtained from Eqs. (\ref{prob_X_QM}) and (\ref{prob_X_W}), respectively, by replacing $S$, $\rho_\psi$ and $T_{\psi}^{\widehat{A}}(X)$ with $N$, $\rho_N$ and $T_{\rho_N}^{\widehat{A}}(X)$, respectively. Moreover easy calculations show that also GPP can be extended to improper mixtures by introducing the same substitutions and replacing $\rho_{\psi_F}$ with $\rho_{N_F}$. This suggests that improper mixtures could actually be considered as generalized pure states in the ESR model, consistently enlarging the mathematical representation of properties and generalized observables by introducing the foregoing substitutions in Eqs. (\ref{prob_X_QM})--(\ref{family_gen_bs}), where $\mathscr V$ must then be intended to denote the set of all density operators on $\mathscr H$.  It is interesting to observe that a similar proposal was made by other authors in a different context \cite{a99}, which supports its reasonableness. Moreover assumption AX can be extended to improper mixtures, which allows one to recover the quantum formalism for mixtures in the ESR model and suggests that QM actually deals only with improper mixtures \cite{bc81,f57,p68,ba98}. But, of course, all the probabilities that occur in the formalism of QM must be interpreted as conditional rather than absolute in the ESR model, at variance with their standard interpretation.\footnote{One could generalize further our treatment by considering more complicate preparation procedures that prepare proper mixtures of improper mixtures, proper mixtures of proper mixtures, etc. But the mathematical formalism needed to describe these procedures becomes fastly more and more complicated, and we avoid this kind of generalization for the sake of simplicity in this paper.}

\section{Testable predictions in the ESR model\label{testable}}
Let us consider a composite physical system made up of two far apart spin--$\frac{1}{2}$ quantum particles prepared in the singlet spin state. We have recently shown that, if one performs spin measurements on each particle, then the ESR model predicts an upper bound for the detection probabilities associated with the spin observables which occurs also in the case of idealized measurements, hence it does not depend on the detection efficiencies of the measuring apparatuses \cite{gs08,gs10}. This result proves that the ESR model is, at least in principle, falsifiable. We aim to discuss in this section another testable prediction that distinguishes the ESR model from QM.

Let the physical system $\Omega$ be a spin $\frac{1}{2}$ quantum particle and let $\Sigma_z$ be the quantum observable ``spin of $\Omega$ along the z--axis''. Let $S_{+}$ and $S_{-}$ be the eigenstates corresponding to the eigenvalues $+1$ and $-1$ of $\Sigma_{z}$, represented by the projection operators $|+\rangle\langle+|$ and $|-\rangle\langle -|$, respectively, and let $M$ be a proper mixture of $S_{+}$ and $S_{-}$ with probabilities $p_{+}$ and $p_{-}=1-p_{+}$, respectively. Then $M$ is represented in the ESR model by the family of pairs
\begin{equation}
\{ ( p_{+} \frac{p_{S_+}^{d}(F)}{p_{M}^{d}(F)} |+\rangle\langle+| + p_{-} \frac{p_{S_-}^{d}(F)}{p_{M}^{d}(F)} |-\rangle\langle -|, p_{M}^{d}(F)) \}_{F \in {\mathcal F}} ,
\end{equation}
with $p_{M}^{d}(F)=p_+p_{S_+}^{d}(F)+p_-p_{S_-}^{d}(F)$. Moreover, let $\Sigma_{n}$ be the quantum observable ``spin of $\Omega$ along the direction $n$'', let $S_n$ be the eigenstate corresponding to the value $+1$ of $\Sigma_{n}$, represented by the projection operator $|+_{n}\rangle\langle +_{n}|$, and let the property $F_{n}=(\Sigma_{n}, \{+1 \})$ be measured on an individual example of $\Omega$ in the state $M$. One can then calculate the conditional probability $p_{M}(F_n)$ and get
\begin{eqnarray} 
p_{M}(F_n)=Tr [ \rho_{M}(F_n)|+_{n}\rangle\langle +_{n}| ]
=Tr [ ( \frac{p_{+}p_{S_+}^{d}(F_n)}{p_+ p_{S_+}^{d}(F_n)+p_- p_{S_-}^{d}(F_n)} |+\rangle\langle +|  \nonumber \\
+ \frac{p_{-}p_{S_-}^{d}(F_n)}{p_+ p_{S_+}^{d}(F_n)+p_- p_{S_-}^{d}(F_n)} |-\rangle\langle -|) |+_{n}\rangle\langle+_{n}|]. \label{conditional}
\end{eqnarray}

The overall probability $p_{M}^{t}(F_n)$ is instead given by
\begin{eqnarray}
p_{M}^{t}(F_n)=p_{M}^{d}(F_n)p_{M}(F_n) \nonumber \\
=Tr [(p_+ p_{S_+}^{d}(F_n)|+\rangle\langle +| 
+p_- p_{S_-}^{d}(F_n)|-\rangle\langle-|)|+_{n}\rangle\langle+_{n}|]. \label{overall}
\end{eqnarray}
Finally, the probability that an example of $\Omega$ in the state $M$ yield the yes outcome when an ideal measurement of $F_n$ is performed on it is given in QM by
\begin{equation}  \label{quantum}
p_{M}^{Q}(F_n)=Tr [(p_+ |+\rangle\langle +|+p_- |-\rangle\langle-|)|+_{n}\rangle\langle+_{n}|].
\end{equation}
By comparing Eqs. (\ref{conditional}), (\ref{overall}) and (\ref{quantum}) we get
\begin{equation}
p_{M}^{Q}(F_n)=p_{M}(F_n)  \ \ \textrm{iff}  \ \  p_{S_+}^{d}(F_n)=p_{S_-}^{d}(F_n) \label{case1} 
\end{equation}
and
\begin{equation}
p_{M}^{Q}(F_n)=p_{M}^{t}(F_n)  \ \  \textrm{iff}  \ \  p_{S_+}^{d}(F_n)=p_{S_-}^{d}(F_n)=1,\label{case2}
\end{equation}
but, in general,
\begin{equation} \label{exp}
p_{M}(F_n) \ne p_{M}^{Q}(F_n) \ne p_{M}^{t}(F_n).
\end{equation}

Equation (\ref{exp}) shows that the predictions of the ESR model do not coincide with the predictions of QM. One can then check Eq. (\ref{exp}) with different choices of $n$ (at least in principle: one should indeed be able to construct measurements that are very close to idealized measurements, \emph{i.e.}, with efficiency close to 1). Should the predictions of QM be violated one would get a clue in favour of the ESR model, and try to determine experimentally the unknown parameters $p_{S_+}^{d}(F_n)$ and $p_{S_-}^{d}(F_n)$, then checking Eqs. (\ref{conditional}) and (\ref{overall}). Should instead the predictions of QM be fulfilled, one must remind that $p_{M}^{Q}(F_n)$ expresses an overall probability. Because of Eq. (\ref{case2}) the obtained result would be compatible with the ESR model only if $p_{M}^{t}(F_n)=p_{M}^{Q}(F_n)$ for all choices of $n$, which betrays the spirit of the ESR model and can be seen as a falsification of it.\footnote{This conclusion may be satisfactory from a theoretical point of view but disappointing for an experimentalist, because actual measuring devices usually have efficiencies much lower than 1. Similar problems occur, however, with Bell's inequalities. For instance, the Clauser--Horne--Shimony--Holt inequality \cite{chsh69} cannot be tested directly because of the low efficiencies of the measuring devices, hence real tests check derived inequalities whose proofs require further problematic assumptions, as fair sampling \cite{b87,sa04,k09}.}


\begin{thebibliography}{99}

\bibitem{b64} Bell J S 1964 \emph{Physics} \textbf{1} 195

\bibitem{b66} Bell J S 1966 \emph{Rev. Mod. Phys.} \textbf{38} 447

\bibitem{ks67} Kochen S and Specker E P 1967 \emph{J. Math. Mech.} \textbf{17} 59

\bibitem{blm91} Busch P, Lahti P J and Mittelstaedt P 1991 \emph{The Quantum Theory of Measurement} (Berlin: Springer)

\bibitem{bs96} Busch P and Shimony A 1996 \emph{Stud. His. Phil. Mod. Phys.} \textbf{27B} 397

\bibitem{b98} Busch P 1998 \emph{Int. J. Theor. Phys.} \textbf{37} 241

\bibitem{bvn36} Birkhoff G and von Neumann J 1935 \emph{Ann. Mat.} \textbf{37} 823

\bibitem{r98} R\'{e}dei N 1998 \emph{Quantum Logic in Algebraic Approach} (Dordrecht: Kluwer)

\bibitem{dcgg04} Dalla Chiara M L, Giuntini R and Greechie R 2004 \emph{Reasoning in Quantum Theory} (Dordrecht: Kluwer)

\bibitem{bc81} Beltrametti E G and Cassinelli G 1981 \emph{The Logic of Quantum Mechanics} (Reading, Mass.: Addison--Wesley)

\bibitem{f65} Feynman R P 1965 \emph{The Character of Physical Laws} (Cambridge, MA: MIT Press)

\bibitem{ga03} Garola C 2003 \emph{Found. Phys. Lett.} \textbf{16} 605

\bibitem{gp04} Garola C and Pykacz J 2004 \emph{Found. Phys.} \textbf{34} 449

\bibitem{gs09} Garola C and Sozzo S 2009 \emph{Europhys. Lett.} \textbf{86} 20009

\bibitem{gs08} Garola C and Sozzo S 2010 \emph{Int. J. Theor. Phys.} \textbf{49} 3101

\bibitem{sg08} Sozzo S and Garola C 2010 \emph{Int. J. Theor. Phys.} \textbf{49}, 3262

\bibitem{gs10} Garola C and Sozzo S 2011 \emph{Found. Phys.} \textbf{41} 424

\bibitem{gs10b} Garola C and Sozzo S 2011 \emph{Int. J. Theor. Phys.} DOI 10.1007/s10773-011-0743-9, in print

\bibitem{gs10c} Garola C and Sozzo S 2011 \emph{Theor. Math. Phys.} \textbf{168}(1) 914 

\bibitem{a2009} Adenier G 2009 in \emph{Foundations of Probability and Physics-5}, Accardi L \emph{et al.} (Eds.) \textbf{1101} 8 (New York: AIP)

\bibitem{des76} d'Espagnat B 1976 \emph{Conceptual Foundations of Quantum Mechanics} (Reading, Mass.: Benjamin)

\bibitem{tb05} Timpson C G and Brown H R 2005 \emph{Int. J. Quant. Inf.} \textbf{3} 679

\bibitem{f57} Fano U 1957 \emph{Rev. Mod. Phys.} \textbf{29} 74

\bibitem{p68} Park J L 1968  \emph{Phil. Sci.} \textbf{35} 205 and 389

\bibitem{ba98} Ballentine L E 1998 \emph{Quantum Mechanics. A Modern Development} (Singapore: World Scientific)

\bibitem{a99} Aerts D 1999 \emph {Int. J. Theor. Phys.} \textbf{38} 289

\bibitem{chsh69} Clauser J F, Horne M A, Shimony A and Holt R A 1969 \emph{Phys. Rev. Lett.} \textbf{23} 880

\bibitem{b87}  Bell J S 1987 \emph{Speakable and Unspeakable in Quantum Mechanics} (Cambridge: Cambridge University Press)

\bibitem{sa04} Santos E 2004 \emph{Found. Phys.} \textbf{34} 1643

\bibitem{k09} Khrennikov A 2009 \emph{Contextual Approach to Quantum Formalism} (Berlin: Springer)

\end{thebibliography}
\end{document}